\documentclass[%
twocolumn,
unsortedaddress,
nofootinbib,
amsmath,
amssymb,
aps,
pre
]{revtex4-2}

\usepackage{graphicx} %,xcolor}% Include figure files
\usepackage{dcolumn}% Align table columns on decimal point
\usepackage{bm}% bold math
\usepackage{wrapfig}
\usepackage{float}
\usepackage[colorlinks=true,linkcolor=blue,citecolor=blue,urlcolor=blue]{hyperref}

\usepackage[dvipsnames]{xcolor}
\usepackage{ulem}

\begin{document}
\title{Theory of associating polymers with annealed and quenched sticker disorder: Mean-field solution and phase behavior}
\author{Sofia Moschin}
\email{smoschin@sissa.it}
\affiliation{
Scuola Internazionale Superiore di Studi Avanzati (SISSA), Via Bonomea 265, 34136 Trieste, Italy
}
\author{Achille Giacometti}
\email{achille.giacometti@unive.it}
\affiliation{
Dipartimento di Scienze Molecolari e Nanosistemi, Universit\`a Ca' Foscari Venezia, 30123 Venezia, Italy
}
\affiliation{
European Centre for Living Technology (ECLT) Ca' Bottacin, 3911 Dorsoduro Calle Crosera, 30123 Venezia, Italy
}
\author{Amos Maritan}
\email{amos.maritan@pd.infn.it}
\affiliation{
Laboratory of Interdisciplinary Physics, Department of Physics and Astronomy ``G. Galilei'', University of Padova, Padova, Italy and INFN, Sezione di Padova, via Marzolo 8, 35131 Padova, Italy
}
\author{Angelo Rosa}
\email{anrosa@sissa.it}
\affiliation{
Scuola Internazionale Superiore di Studi Avanzati (SISSA), Via Bonomea 265, 34136 Trieste, Italy
}
\date{\today}
%

%%%
\begin{abstract}
%%%
We develop a density-functional theory for solutions of associating polymers where attractions among charged monomers (stickers) are represented by local binary degrees of freedom, which are randomly placed along the chains.
Extending the original Garel and Orland's field-theoretic scheme for single-chain systems to an ensemble of interacting chains, we give the exact formulations of the model in both cases of annealed and quenched distributions of charges which we solve at the mean-field level.
The solution produces qualitatively different free energy functionals.
In the annealed case, the theory naturally yields a nontrivial scalar order parameter for the fraction of bonded sticker monomers and a self-consistent mass-action law at the saddle point.
By contrast, in the quenched case no independent bonding order parameter emerges and the main effect is a renormalization of the effective two-body and three-body interaction parameters.
The formulation is microscopic at the Hamiltonian level and, within the same field-theoretic framework, provides a systematic starting point for fluctuation corrections beyond the mean-field approximation.
%%%
\end{abstract}
%%%

%%%
\maketitle
%%%

%%%
\section{Introduction}\label{sec:Intro}
%%%
Thermoreversible association between sparse attractive groups embedded along polymer chains provides a common physical mechanism for controlling the slow down of chain dynamics until its complete arrest via gelation in synthetic solutions~\cite{ZhangColby2018,Golkaram2019,Cai-PhysRevLett2023} and for promoting phase separation in multivalent biomacromolecular systems~\cite{Brangwynne2009_Science_PGranules,HymanWeberJulicher2014_AnnuRevCellDevBiol,Banani2017_NatRevMCB_Condensates,AlbertiHyman2021_NatRevMCB_StressAging,LyonPeeplesRosen2021_NatRevMCB_Framework,PappuCohenDarFaragKar2023_ChemRev}. 
A recurring situation is that where many proteins can undergo liquid-liquid phase separation, forming liquid-like condensed phases that act as biomolecular condensates in cells~\cite{FuxreiterVendruscolo2020_Cell}.

Classical theories established the mean-field picture of reversible network formation and coexistence~\cite{Flory1941_JACS_Gelation,Stockmayer1943_JCP_GelFormation}.
A notable example of this approach is provided by the Semenov-Rubinstein description of associating polymers where specific, saturable interaction motifs ({\it stickers}) are placed {\it periodically} along the chain backbone separated by monomer beads ({\it spacers}) with excluded volume only~\cite{SemenovRubinstein1998_Statics,RubinsteinSemenov1998_Dynamics,RubinsteinDobrynin1999_GelsReview}.
With respect to spacers, stickers can associate in pairs reversibly whenever the associated energy overcomes thermal fluctuations.
This pioneering study generated further analysis, some of which very recently~\cite{Danielsen2023phase,Chen2025sol,Cappa2026single}.
The physical picture emerging from all these studies hinges upon the competition between phase separation, characteristic of a gas of polymers, and the association and formation of a reversible network as promoted by the presence of the stickers.

While highly successful, the Semenov-Rubinstein theory is restricted to polymers with regularly spaced stickers.
In contrast, biomolecular condensates often involve intrinsically heterogeneous macromolecules whose interaction motifs are distributed nonuniformly along the sequence.
Recent studies have emphasized the importance of sticker valence, spacing, and interaction strength in determining phase behavior~\cite{ChoiPappu2019_BiophysJ_Abstract,ChoiHolehousePappu2020_AnnuRevBiophys,Harmon2017_eLife_Linkers,Wang2018_Cell_Grammar,Martin2020_Science_PLCD,MittagPappu2022_MolCell_Framework,RuffPappuHolehouse2019_COSB}.
These developments motivate theoretical approaches capable of describing sequence-disordered sticker architectures in dense polymer solutions, where phase coexistence emerges as a collective thermodynamic phenomenon.

Beyond biomolecular systems, sequence disorder can also be realized and controlled experimentally in DNA-mediated associating materials~\cite{Biffi2013_PNAS_DNAnanostars}.
In such systems, sticker degrees of freedom may be effectively {\it quenched}, corresponding to fixed interaction architectures, or {\it annealed} through reversible activation and deactivation mechanisms.
This distinction provides a unique opportunity to investigate experimentally how different realizations of disorder influence reversible network formation and phase separation, thereby motivating a unified theoretical framework capable of treating both cases on equal footing.

In this work, we address these issues by focusing on the role of disorder in sticker distributions, as suggested by experimental observations.
To this end, the original Semenov-Rubinstein combinatorial approach is reformulated within a microscopic field-theoretical framework.
This formulation enables a controlled treatment of different types of disorder (annealed and quenched) and provides a systematic route for further refinements.
Building on the theoretical work of Garel and Orland on model biopolymers with contact interactions~\cite{GarelOrland_1988a,GarelOrland_1988b,GarelOrlandThirumalai1996_AnalyticalTheoriesProteinFolding}, we extend their approach to solutions of interacting flexible polymers containing randomly distributed stickers separated by spacer segments.
Within this unified framework, we formulate a density-functional field theory for associating polymers with sequence-disordered sticker distributions and investigate the effect of annealed and quenched disorder on the mean-field free energy.
The resulting theory establishes a microscopic connection between density-functional polymer field theory and the classical Semenov-Rubinstein description, while generalizing the latter to random sticker architectures relevant to biomolecular condensates.

The paper is structured as it follows.
In Sec.~\ref{sec:TheoryModelMethods}, we introduce the polymer model and the field-theoretic methods (complemented, in Appendix~\ref{sec:feynman}, by a formulation for the propagator or the end-to-end spatial distribution of a polymer chain in an external field) to calculate the {\it exact} free energies of the systems for the annealed and quenched distributions of stickers along the chains.
Then, using the saddle point approximation, we obtain mean-field expressions for the free energies.
In particular, as shown in Appendix~\ref{sec:Relation2SR}, the mean-field free energy of the annealed model is related to a generalized formulation of the classical theory {\it \`a la} Semenov and Rubinstein.
In Sec.~\ref{sec:PhaseDiagramNumerics}, we provide a brief summary of the numerical methods for constructing the binodal coexistence curves from the mean-field free energy profiles.
Then, results are presented in Sec.~\ref{sec:Results}.
Finally, we discuss and conclude our work, highlighting in particular future perspectives, in Sec.~\ref{sec:DiscConcls}.

%%%
\section{Theory: model and methods}\label{sec:TheoryModelMethods}
%%%

%%%
\subsection{Polymer model}\label{sec:PolymerModel}
%%%

%%%
\subsubsection{Definition}\label{sec:PolymerModel-Definition}
%%%
We consider a system of $K$ polymer chains (labels $\alpha=1,\dots,K$), where each chain consists of $N$ monomers (labels $i=1,\dots,N$) whose spatial coordinates are denoted by ${\vec r}_i^{\, (\alpha)}$.
Each monomer contains an additional ``binary'' degree of freedom, the ``sticker'' variable denoted by $q_i^{(\alpha)}$, which can take one of two possible values ``$0$'' or ``$q$'' depending on the prescription which will be specified later (see Sec.~\ref{sec:annealed-vs-quenched-disorder}).
Then, with $\beta = 1/(k_BT)$ the usual Boltzmann factor, the total Hamiltonian of the system
\begin{eqnarray}\label{eq:TotalHamiltonian}
\beta {\mathcal H}(\{ {\vec r}_i^{\, (\alpha)}, q_i^{(\alpha)} \}) & = &
\beta {\mathcal H}_{\rm conn}(\{ {\vec r}_i^{\, (\alpha)} \}) +
\beta {\mathcal H}_{\rm vol}(\{ {\vec r}_i^{\, (\alpha)} \}) \nonumber\\
& & + \beta {\mathcal H}_{\rm stick}(\{ {\vec r}_i^{\, (\alpha)}, q_i^{(\alpha)} \})
\end{eqnarray}
consists of the following three terms:
\begin{enumerate}
\item {\it Chain connectivity}, $\beta {\mathcal H}_{\rm conn}$ --
This term encodes the connectivity constraints of polymer chains and accounts for the entropic elasticity associated with chain conformations, therefore obeying Gaussian statistics~\cite{DoiEdwards,RubinsteinColby2003_PolymerPhysics},
\begin{equation}\label{eq:H_conn_discrete}
\beta {\mathcal H}_{\rm conn}(\{ {\vec r}_i^{\, (\alpha)} \}) = \frac3{2 b^2} \sum_{\alpha=1}^K \sum_{i=1}^{N-1} \left( {\vec r}_{i+1}^{\, (\alpha)} - {\vec r}_i^{\, (\alpha)} \right)^2 \, ,
\end{equation}
where $b$ is the monomer-monomer distance along the chain.
\item {\it Volume interactions}, $\beta {\mathcal H}_{\rm vol}$ --
Volume interactions between monomers are described as the sum of generic, two- and three-body terms, {\it i.e.}
\begin{align}\label{eq:Hev_micro}
& \beta {\mathcal H}_{\rm vol}(\{ {\vec r}_i^{\, (\alpha)} \}) = \frac{w_2}2 \sum_{\alpha,\beta=1}^K \sum_{i,j=1}^N f_2 \! \left( {\vec r}_i^{\, (\alpha)}, {\vec r}_j^{\, (\beta)} \right) \nonumber\\
& + \frac{w_3}6 \sum_{\alpha,\beta,\gamma=1}^K \sum_{i,j,k=1}^N f_3 \! \left( {\vec r}_i^{\, (\alpha)}, {\vec r}_j^{\, (\beta)}, {\vec r}_k^{\, (\gamma)} \right) \, , \nonumber\\
\end{align} 
where $w_2$ and $w_3$, equivalent to the second and third virial coefficients~\cite{RubinsteinColby2003_PolymerPhysics,Bhattacharjee2013flory}, define the strengths of the interactions:
$w_2$ depends on the temperature and may be positive or negative (depending on whether the polymers are in good or in poor solvent~\cite{RubinsteinColby2003_PolymerPhysics} conditions) while $w_3$ is, generally, positive.
Then, the functions $f_2$ and $f_3$:
(i)
are symmetric with respect to permutations of monomers' coordinates, {\it i.e.} $f_2 \! \left( {\vec r}_i^{\, (\alpha)}, {\vec r}_j^{\, (\beta)} \right) = f_2 \! \left( {\vec r}_j^{\, (\beta)}, {\vec r}_i^{\, (\alpha)} \right)$ and $f_3 \! \left( {\vec r}_i^{\, (\alpha)}, {\vec r}_j^{\, (\beta)}, {\vec r}_k^{\, (\gamma)} \right) = f_3 \! \left( {\vec r}_j^{\, (\beta)}, {\vec r}_i^{\, (\alpha)}, {\vec r}_k^{\, (\gamma)} \right)$ and so on;
(ii)
depend only on monomers' distance vectors, {\it i.e.} $f_2({\vec x}, {\vec y}) = f_2(\vec 0, {\vec y} - {\vec x})$ and $f_3({\vec x}, {\vec y}, {\vec z}) = f_3(\vec 0, {\vec y} - {\vec x}, {\vec z} - {\vec x})$;
(iii)
are short-ranged (typically, of the order of monomer size $\approx b$, see Eq.~\eqref{eq:H_conn_discrete}).
The following simplifying approximations can then be used:
\begin{eqnarray}
1 & = & \int {\rm d}{\vec x} \, f_2(\vec 0, {\vec x}) \approx b^3 f_2(\vec 0, \vec 0) \, , \label{eq:Integratef2} \\
1 & = & \int {\rm d}{\vec x} \, {\rm d}{\vec y} \, f_3(\vec 0, {\vec x}, {\vec y}) \approx b^6 f_3(\vec 0, \vec 0, \vec 0) \, . \label{eq:Integratef3}
\end{eqnarray}
\item {\it Sticker-sticker interactions}, $\beta {\mathcal H}_{\rm stick}$ --
The sticker Hamiltonian has a form similar to $\beta {\mathcal H}_{\rm vol}$:
\begin{align}\label{eq:Hstick_micro}
& \beta {\mathcal H}_{\rm stick}(\{ {\vec r}_i^{\, (\alpha)}, q_i^{(\alpha)} \}) = - \frac{w_2^{\rm stick}}2 \sum_{\alpha,\beta=1}^K \sum_{i,j=1}^N q_i^{(\alpha)} q_j^{(\beta)} f_2 \! \left( {\vec r}_i^{\, (\alpha)}, {\vec r}_j^{\, (\beta)} \right) \nonumber\\
& + \frac{w_3^{\rm stick}}6 \sum_{\alpha,\beta,\gamma=1}^K \sum_{i,j,k=1}^N q_i^{(\alpha)} q_j^{(\beta)} q_k^{(\gamma)} f_3 \! \left( {\vec r}_i^{\, (\alpha)}, {\vec r}_j^{\, (\beta)}, {\vec r}_k^{\, (\gamma)} \right) \, .
\end{align}
Taking $w_2^{\rm stick} > 0$, the two-body terms controls the effective short-range attraction between stickers, while the cubic term with coupling $w_3^{\rm stick} > 0$ penalizes multiple-sticker localization and ensures thermodynamic stability.
\end{enumerate}
%

%%%
\subsubsection{Formulating the model in terms of monomer- and sticker-density fields}\label{sec:ExpressingModel_with_Densities}
%%%
It is convenient to reformulate the model in terms of the following monomer- and sticker-density fields,
\begin{eqnarray}
\rho(\vec r) & = & \sum_{\alpha=1}^K \sum_{i=1}^N  \, \delta \! \left( {\vec r} - {\vec r}_i^{\, (\alpha)} \right) \, , \label{eq:rho_micro} \\
Q(\vec r) & = & \sum_{\alpha=1}^K \sum_{i=1}^N \, q_i^{(\alpha)} \, \delta \! \left( \vec r - \vec r_i^{\, (\alpha)} \right) \, , \label{eq:Q_micro}
\end{eqnarray}
whose integrals over space, $\int {\rm d}\vec r \, \rho(\vec r)$ and $\int {\rm d}\vec r \, Q(\vec r)$, give the total number of monomers ($=KN$) and stickers of the system, respectively.
Using the definitions~\eqref{eq:rho_micro} and~\eqref{eq:Q_micro} and the approximations~\eqref{eq:Integratef2} and~\eqref{eq:Integratef3}, the terms $\beta {\mathcal H}_{\rm vol}$ (Eq.~\eqref{eq:Hev_micro}) and $\beta {\mathcal H}_{\rm stick}$ (Eq.~\eqref{eq:Hstick_micro}) can be expressed as the following {\it functionals} of $\rho(\vec r)$ and $Q(\vec r)$, respectively:\footnote{Notice that this is true up to additive self-interaction terms which, however, produce merely a global shift of the free energy reference level and are, therefore, irrelevant.}
\begin{align}\label{eq:Hev_micro_density}
& \beta {\mathcal H}_{\rm vol} = \beta {\mathcal H}_{\rm vol} [\rho] = \nonumber\\
& \frac{w_2}2 \int {\rm d}\vec r \, {\rm d}{\vec r}\,' \rho(\vec r) \, \rho({\vec r}\,') \, f_2(\vec 0, {\vec  r}\,' - \vec r) \nonumber\\
& + \frac{w_3}6 \int {\rm d}\vec r \, {\rm d}{\vec r}\,' \, {\rm d}{\vec r}\,'' \rho(\vec r) \, \rho({\vec r}\,') \, \rho({\vec r}\,'') \, f_3(\vec 0, {\vec r}\,' - {\vec r}, {\vec r}\,'' - {\vec r}) \nonumber\\
& \approx \frac{w_2}2 \int {\rm d}\vec r \, \rho(\vec r)^2 + \frac{w_3}6 \int {\rm d}\vec r \, \rho(\vec r)^3 \, ,
\end{align}
and
\begin{align}\label{eq:Hstick_micro_density}
& \beta {\mathcal H}_{\rm stick} = \beta {\mathcal H}_{\rm stick} [Q] = \nonumber\\
& -\frac{w_2^{\rm stick}}2 \int {\rm d}\vec r \, {\rm d}{\vec r}\,' Q(\vec r) \, Q({\vec r}\,') \, f_2(\vec 0, {\vec  r}\,' - \vec r) \nonumber\\
& + \frac{w_3^{\rm stick}}6 \int {\rm d}\vec r \, {\rm d}{\vec r}\,' \, {\rm d}{\vec r}\,'' Q(\vec r) \, Q({\vec r}\,') \, Q({\vec r}\,'') \, f_3(\vec 0, {\vec r}\,' - {\vec r}, {\vec r}\,'' - {\vec r}) \nonumber\\
& \approx -\frac{w_2^{\rm stick}}2 \int {\rm d}\vec r \, Q(\vec r)^2 + \frac{w_3^{\rm stick}}6 \int {\rm d}\vec r \, Q(\vec r)^3 \, .
\end{align}
Eqs.~\eqref{eq:Hev_micro_density} and~\eqref{eq:Hstick_micro_density} constitute the starting point of our theory.

%%%
\subsection{Averaging out the effect of stickers distribution: annealed {\it vs.} quenched model}\label{sec:annealed-vs-quenched-disorder}
%%%

%%%
\subsubsection{Annealed model}\label{sec:annealed}
%%%
Assume that the stickers are mobile with a relaxation time comparable to the Rouse relaxation time~\cite{DoiEdwards,RubinsteinColby2003_PolymerPhysics} of each single chain.
Then the disorder is {\it annealed} and has to be treated on equal foot as all other degrees of freedom.
For a {\it given} realization of the sticker variables $\{q_i^{(\alpha)}\}$, the Hamiltonian Eq.~\eqref{eq:TotalHamiltonian} gives the following partition function:
\begin{equation}\label{eq:Z-micro}
Z(\{ q_i^{\alpha} \}) = \int \prod_{\alpha=1}^K \prod_{i=1}^N {\rm d} \vec r_i^{\,(\alpha)} \exp\!\left[ -\beta {\mathcal H} ( \{ \vec r_i^{\,\alpha}, q_i^{\,\alpha} \} ) \right] \, .
\end{equation}
Then, assuming that the sticker variables $\{ q_i^{(\alpha)} \}$ are randomly assigned according to some probability distribution $P(\{ q_i^{(\alpha)} \})$, the average partition function $\overline Z$ of the system is
\begin{equation}\label{partition}
e^{-\beta F_{\rm A}} = \overline{Z} 
= \frac1{K!} \int \prod_{\alpha=1}^K \prod_{i=1}^N {\rm d}q_i^{(\alpha)} P(\{ q_i^{(\alpha)} \}) \, Z(\{ q_i^{(\alpha)} \}) \, ,
\end{equation}
where $F_{\rm A}$ is the corresponding annealed free energy of the system and the prefactor $1/K!$ accounts for the indistinguishability of the different chains.

Eq.~\eqref{partition} -- which defines completely the thermodynamics of the {\it annealed} model -- can be usefully reformulated in terms of the collective fields $\rho(\vec r)$ and $Q(\vec r)$ (Eqs.~\eqref{eq:rho_micro} and~\eqref{eq:Q_micro}, respectively) using the following standard identities:\footnote{It is tacitly assumed that the corresponding measures (${\mathcal D} \rho$, ${\mathcal D} \phi$, etc.) of the functional integrals include all proper normalization constants.}
\begin{eqnarray}\label{eq:rho_constraint_continuum}
1 & = &
\int {\mathcal D}\rho \, \delta\!\left[ \rho(\vec r) - \sum_{\alpha,i} \, \delta(\vec r - \vec r_i^{\,(\alpha)}) \right] \nonumber\\
& = & \int {\mathcal D}\rho \, {\mathcal D}\phi \, \exp\!\left[ {\rm i} \int {\rm d}{\vec r} \, \phi(\vec r) \left( \rho(\vec r) - \sum_{\alpha,i} \, \delta(\vec r - \vec r_i^{\,(\alpha)}) \right) \right] \nonumber\\
& = & \int {\mathcal D}\rho \, {\mathcal D}\phi \, \exp\!\left[ {\rm i} \int {\rm d}{\vec r} \, \phi(\vec r) \rho(\vec r) - {\rm i} \sum_{\alpha,i} \, \phi(\vec r_i^{\,(\alpha)}) \right] \, ,
\end{eqnarray}
and
\begin{eqnarray}\label{eq:Q_constraint_continuum}
1 & = &
\int {\mathcal D}Q \, \delta\!\left[ Q(\vec r) - \sum_{\alpha,i} \, q_i^{(\alpha)}\, \delta(\vec r - \vec r_i^{\,(\alpha)}) \right] \nonumber\\
& = & \int {\mathcal D}Q \, {\mathcal D}\lambda \, \exp\!\left[ {\rm i} \int {\rm d}{\vec r} \, \lambda(\vec r) \, \left( Q(\vec r) - \sum_{\alpha,i} \, q_i^{(\alpha)} \, \delta(\vec r - \vec r_i^{\,(\alpha)}) \right) \right] \nonumber\\
& = & \int {\mathcal D}Q \, {\mathcal D}\lambda \, \exp\!\left[ {\rm i} \int {\rm d}{\vec r} \, \lambda(\vec r) \, Q(\vec r) - {\rm i} \sum_{\alpha,i} \, q_i^{(\alpha)} \, \lambda (\vec r_i^{\,(\alpha)}) \right] 
\end{eqnarray}
where $\phi(\vec r_i^{\,(\alpha)})$ (in Eq.~\eqref{eq:rho_constraint_continuum}) and $\lambda(\vec r_i^{\,(\alpha)})$ (in Eq.~\eqref{eq:Q_constraint_continuum}) are real auxiliary fields.
Using identities~\eqref{eq:rho_constraint_continuum} and~\eqref{eq:Q_constraint_continuum} and the definitions of Sec.~\ref{sec:PolymerModel}, the partition function $Z(\{ q_i^{\alpha} \})$ (Eq.~\eqref{eq:Z-micro}) can be rearranged as the following:
\begin{widetext}
\begin{eqnarray}\label{eq:Z-micro-bis}
Z(\{ q_i^{\alpha} \})
& = & \int {\mathcal D}\rho \, {\mathcal D}\phi \, {\mathcal D}Q \, {\mathcal D}\lambda \, \exp\left[ \int {\rm d}\vec r \left( {\rm i} \phi(\vec r)\rho(\vec r) - \frac{w_2}2 \rho(\vec r)^2 - \frac{w_3}6 \rho(\vec r)^3 \right) \right] \nonumber\\
& & \times \exp\left[ \int {\rm d}\vec r \left( {\rm i} \lambda(\vec r)Q(\vec r) + \frac{w_2^{\rm stick}}2 Q(\vec r)^2 - \frac{w_3^{\rm stick}}6 Q(\vec r)^3 \right) \right] \nonumber\\
& & \int \prod_{\alpha=1}^K \prod_{i=1}^N {\rm d} \vec r_i^{\,(\alpha)} \exp\!\left[ -\beta {\mathcal H}_{\rm conn} ( \{ \vec r_i^{\,\alpha} \} ) \right] \, \exp \! \left[ - {\rm i} \sum_{\alpha,i} \, \phi(\vec r_i^{\,(\alpha)}) \right] \,  \exp \! \left[- {\rm i} \sum_{\alpha,i} \, q_i^{(\alpha)}\lambda(\vec r_i^{\,(\alpha)}) \right] \, .
\end{eqnarray}
\end{widetext}
In particular, assuming that the sticker variables are independent random variables drawn from the Bernoulli-like distribution:
\begin{equation}\label{eq:Pq_bernoulli}
P(q_i^{(\alpha)}) = (1-c) \, \delta(q_i^{(\alpha)}) + c\,\delta(q_i^{(\alpha)}-q) \, , \,\,\, 0<c<1 \, ,
\end{equation}
where $q>0$ is the sticker ``charge'' and $c$ is the (bare) fraction of sticker-bearing monomers, the partition function $\overline{Z}$ (Eq.~\eqref{partition}) of the annealed model is immediately obtained from Eq.~\eqref{eq:Z-micro-bis} and the simple relation
\begin{equation}\label{eq:q_average}
\int {\rm d}q' \, P(q') \, e^{ -{\rm i} q' \lambda(\vec r) } = (1-c) + c\,e^{ - {\rm i} q \lambda(\vec r) } \, ,
\end{equation}
with the result
\begin{widetext}
\begin{equation}\label{eq:Z_DFT_continuum}
\overline{Z} = \frac1{K!} \int \mathcal D\rho\,\mathcal D\phi\,\mathcal DQ\,\mathcal D\lambda\; \exp\!\left[ {\mathcal G}_{\rm vol}[\rho, \phi] + {\mathcal G}_{\rm stick}[Q, \lambda] + K \ln(\zeta[\phi, \lambda]) \right] \, ,
\end{equation}
\end{widetext}
where:
\begin{widetext}
\begin{eqnarray}
\mathcal G_{\rm vol}[\rho, \phi] & = & \int {\rm d} \vec r \left[ {\rm i} \phi(\vec r) \rho(\vec r) - \frac{w_2}2 \rho(\vec r)^2 - \frac{w_3}6 \rho(\vec r)^3 \right] \, , \label{eq:S30} \\
\mathcal G_{\rm stick}[Q,\lambda] & = & \int {\rm d} \vec r \left[ {\rm i} \lambda(\vec r) Q(\vec r) + \frac{w_2^{\rm stick}}2 Q(\vec r)^2 - \frac{w_3^{\rm stick}}6 Q(\vec r)^3 \right] \, , \label{eq:S31}
\end{eqnarray}
\end{widetext}
and
\begin{widetext}
\begin{equation}\label{eq:zeta_phi_lambda_cont}
\zeta[\phi,\lambda] = \int \prod_{i=1}^N {\rm d} \vec r_i \, \exp\!\left[ -\frac3{2b^2} \sum_{i=1}^{N-1} (\vec r_{i+1} - \vec r_i)^2 - {\rm i} \sum_{i=1}^N \phi(\vec r_i) + \sum_{i=1}^N \ln\!\left( (1-c) + c \, e^{-{\rm i} q\lambda(\vec r_i)} \right) \right] \, ,
\end{equation}
\end{widetext}
which is the partition function for a single Gaussian chain in the ``external'' fields $\phi(\vec r)$ and $\lambda(\vec r)$.

%%%
\subsubsection{Quenched model}\label{sec:quenched_continuum}
%%%
In the quenched model the sticker variables $\{q_i^{(\alpha)}\}$ have relaxation times much longer compared with the Rouse relaxation time~\cite{DoiEdwards,RubinsteinColby2003_PolymerPhysics} of a single chain, and hence are {\it frozen in}.
This requires the average of the free energy rather than the average of the partition function.
To this aim, the usual replica trick~\cite{Edwards1975_JPhysF_SpinGlasses}
\begin{equation}\label{eq:replica_trick}
-\beta F_{\rm Q} = \overline{\ln(Z)} = \lim_{n\to0}\frac{\overline{Z^n}-1}{n} \, ,
\end{equation}
can be exploited, where $\overline{(\cdots)} = \int\prod_{\alpha,i}dq_i^{(\alpha)} \, P(q_i^{(\alpha)})(\cdots)$ denotes the average over the same distribution $P$ (Eq.~\eqref{eq:Pq_bernoulli}) as described in Sec.~\ref{sec:annealed}.
Then, analogously to the annealed case, we introduce the replica-counterparts of the functional constraints Eqs.~\eqref{eq:rho_micro} and~\eqref{eq:Q_micro}.
Namely, for each replica $a=1,\dots,n$, we define
\begin{eqnarray}
\rho^{(a)}(\vec r) & = & \sum_{\alpha=1}^K \sum_{i=1}^N \delta(\vec r - \vec r_i^{\,(\alpha,a)}) \, , \label{eq:rep-rho} \\
Q^{(a)}(\vec r) & = & \sum_{\alpha=1}^K \sum_{i=1}^N q_i^{(\alpha)} \delta(\vec r - \vec r_i^{\,(\alpha,a)}) \, , \label{eq:rep-Q}
\label{eq:rep}
\end{eqnarray}
and introduce corresponding $\delta$-functionals as in Eqs.~\eqref{eq:rho_constraint_continuum} and~\eqref{eq:Q_constraint_continuum} with conjugate fields $\phi^{(a)}$ and $\lambda^{(a)}$. 
After inserting the constraints, the replica-partition function reads:
\begin{widetext}
\begin{equation}\label{eq:S72}
\overline{Z^n} = \frac1{(K!)^n} \int \prod_{a=1}^n {\mathcal D}\rho^{(a)} \, {\mathcal D}\phi^{(a)} \, {\mathcal D}Q^{(a)} \, {\mathcal D}\lambda^{(a)} \;
\exp\!\left\{
\sum_{a=1}^n \Big[ {\mathcal G}_{\rm vol}[\rho^{(a)}, \phi^{(a)}] + {\mathcal G}_{\rm stick}[Q^{(a)}, \lambda^{(a)}] \Big]
\right\}
\Big[ \zeta_n[\{\phi^{(a)}\}, \{\lambda^{(a)}\}] \Big]^K \, ,
\end{equation}
\end{widetext}
where $\mathcal G_{\rm vol}$ and $\mathcal G_{\rm stick}$ have the same functional forms Eqs.~\eqref{eq:S30} and~\eqref{eq:S31} of the annealed model, while for a fixed monomer site $(\alpha,i)$ the same quenched variable $q_i^{(\alpha)}$ enters all replicas, thus producing the local factor
\begin{align}\label{eq:quenched_local_factor_rewrite}
& \overline{ \exp\!\left[- {\rm i} \, q_i^{(\alpha)} \sum_{a=1}^n \lambda^{(a)}(\vec r_i^{\,(\alpha,a)}) \right] } \nonumber\\
& = (1-c) + c \, \exp\!\left[ - {\rm i} \, q\sum_{a=1}^n \lambda^{(a)}(\vec r_i^{\,(\alpha,a)}) \right] \, .
\end{align}
Finally, collecting the connectivity weights given by an analogous sum of harmonic terms as in Eq.~\eqref{eq:H_conn_discrete} and the external potentials, the single-chain contribution of the replicas assumes the form:
\begin{widetext}
\begin{align}\label{eq:S75}
\zeta_n[\{\phi^{(a)}\}, \{\lambda^{(a)}\}]
= \int \prod_{a=1}^n \prod_{i=1}^N {\rm d} \vec r_i^{\,(a)}
& \exp\!\left[ -\frac3{2b^2} \sum_{a=1}^n\sum_{i=1}^{N-1} (\vec r_{i+1}^{\,(a)} - \vec r_i^{\,(a)})^2 - {\rm i} \sum_{a=1}^n\sum_{i=1}^N \phi^{(a)}(\vec r_i^{\,(a)}) \right. \nonumber\\
& \left. + \sum_{i=1}^N \ln\!\Big( (1-c)+c\,e^{- {\rm i} q \sum_{a=1}^n \lambda^{(a)}(\vec r_i^{\,(a)})} \Big) \right] \, .
\end{align}
\end{widetext}
Eq.~\eqref{eq:S75} together with Eq.~\eqref{eq:quenched_local_factor_rewrite} shows that, upon averaging over the charge distribution $P$, distinct replicas couple to each other in a non-trivial manner, in particular the corresponding potential does not simply factorize into single-replica contributions. 

%%%
\subsection{Saddle point approximation}\label{sec:SPapproxAnnealedQuenched}
%%%
Neither the annealed nor the quenched models can be solved exactly.
However, they can be both solved within a mean-field approximation, as we explain in Sec.~\ref{sec:SPapproxAnnealed} (annealed model) and in Sec.~\ref{sec:SPapproxQuenched} (quenched model).
 
%%%
\subsubsection{Annealed model}\label{sec:SPapproxAnnealed}
%%%
Using the saddle point approximation reported in Appendix~\ref{sec:feynman}, the partition function of the annealed model Eq.~\eqref{eq:Z_DFT_continuum} takes the form
\begin{eqnarray}\label{eq:AnnealedforFeynmanKac}
\overline Z
& = & \frac1{K!} \int \mathcal D\rho\,\mathcal D\phi\,\mathcal DQ\,\mathcal D\lambda\;
\exp\!\left[ \mathcal G_{\rm vol}[\rho, \phi] + \mathcal G_{\rm stick}[Q, \lambda] \right] \nonumber\\
& & \times \prod_{\alpha=1}^K \left[ \int {\rm d} \vec r_\alpha \, {\rm d} \vec r_{\alpha,0} \; G_{{\mathcal V}_{\phi,\lambda}}(\vec r_{\alpha,N}, N | \vec r_{\alpha,0}, 0) \right] \, ,
\end{eqnarray}
with the 
``potential'',
\begin{equation}\label{eq:Vphilambda_here}
{\mathcal V}_{\phi,\lambda}(\vec r) = {\rm i} \, \phi(\vec r) -\ln\!\left[ (1-c) + c \, e^{- {\rm i} q \lambda(\vec r)} \right] \, ,
\end{equation}
encoding for the annealed sticker average.
Assuming translational invariance, we restrict to homogeneous fields independent of spatial coordinates, {\it i.e.}
\begin{equation}\label{eq:hom_fields_assumption}
\rho(\vec r) = \rho \, , \,\,\,
\phi(\vec r) = \phi \, , \,\,\,
Q(\vec r) = Q \, , \,\,\,
\lambda(\vec r) = \lambda \, ,
\end{equation}
thus leading to the following forms for the functionals $\mathcal G_{\rm vol}$ and $\mathcal G_{\rm stick}$ (see Eqs.~\eqref{eq:S30} and~\eqref{eq:S31}):
\begin{eqnarray}
\mathcal G_{\rm vol}(\rho, \phi) & = & V \left( {\rm i} \phi\rho - \frac{w_2}2 \rho^2 - \frac{w_3}6 \rho^3 \right) \, , \label{eq:G0_hom_here}\\
\mathcal G_{\rm stick}(Q, \lambda) & = & V \left( {\rm i} \lambda Q + \frac{w_2^{\rm stick}}2 Q^2 - \frac{w_3^{\rm stick}}6 Q^3 \right) \, , \label{eq:GQ_hom_here}
\end{eqnarray}
where $V$ is the volume of the system.
Moreover, the potential~\eqref{eq:Vphilambda_here} becomes constant ${\mathcal V}_{\phi, \lambda} = {\rm i} \, \phi - \ln\!\left[ (1-c) + c\,e^{- {\rm i} \lambda q} \right]$ and, therefore, we can use Eq.~\eqref{eq:time}.
Taken all together and using the Stirling approximation $K! \simeq (K/e)^K$, the free energy of the system per unit volume is given by:
\begin{widetext}
\begin{eqnarray}\label{eq:lnZ_hom_annealed}
\frac{\beta F_{\rm A}}V 
& \equiv & -\frac{\ln(\overline Z)}V \nonumber\\
& = & \frac{K}V\ln\!\left( \frac{K}{eV} \right) - {\rm i} \phi\!\left( \rho-\frac{KN}V \right) + \frac{w_2}2 \rho^2 + \frac{w_3}6 \rho^3 - {\rm i} \lambda Q - \frac{w_2^{\rm stick}}2 Q^2 + \frac{w_3^{\rm stick}}6 Q^3 - \frac{KN}V \ln\!\left[ (1-c) + c\,e^{-{\rm i} \lambda q} \right] \, . \nonumber\\
\end{eqnarray}
\end{widetext}

Minimization of the free energy density Eq.~\eqref{eq:lnZ_hom_annealed} with respect to $\phi$ leads to
\begin{equation}\label{eq:Annealed-SaddlePointCondition-1}
\partial_{\phi}(\beta F_{\rm A} / V) = 0 \Rightarrow \frac{K N}V = \rho \, ,
\end{equation}
a result which could be expected on physical ground.
On the other hand, upon minimization with respect to $\lambda$ and $Q$ leads, one obtains
\begin{eqnarray}
\partial_{\lambda}(\beta F_{\rm A} / V) = 0 & \Rightarrow & Q = \frac{q K N}V \frac{c \, e^{-{\rm i} \lambda q}}{(1-c) + c \, e^{-{\rm i} \lambda q}} \equiv q \pi \rho \, , \nonumber\\
\label{eq:Annealed-SaddlePointCondition-3} \\
\partial_{Q}(\beta F_{\rm A} / V) = 0 & \Rightarrow & -{\rm i} \lambda = w_2^{\rm stick}Q - \frac{w_3^{\rm stick}}2 Q^2 \, , \label{eq:Annealed-SaddlePointCondition-4}
\end{eqnarray}
where we have introduced the (annealed) ``dressed'' probability for a monomer to be a sticker:
\begin{equation}\label{eq:OccBindingProb}
\pi \equiv \frac{ c \, e^{-{\rm i} \lambda q} }{ (1-c) + c \, e^{-{\rm i} \lambda q} } \, .
\end{equation}
Finally, the annealed free energy~\eqref{eq:lnZ_hom_annealed} as a function of monomer density $\rho$ takes the form:
\begin{widetext}
\begin{equation}\label{eq:F_A_final_correct}
\frac{\beta F_{\rm A}}V = \frac{\rho}N \ln\!\left( \frac{\rho}{eN} \right) + \frac{w_2}2 \rho^2 + \frac{w_3}6 \rho^3 - \frac{w_2^{\rm stick}}2 q^2 \pi^2 \rho^2 + \frac{w_3^{\rm stick}}6 q^3 \pi^3 \rho^3 + \rho \left[ \pi \ln\!\left( \frac{\pi}c \right) + (1-\pi) \ln\!\left( \frac{1-\pi}{1-c} \right) \right] \, ,
\end{equation}
\end{widetext}
with the following {\it mass-action}-like relation (derived from Eq.~\eqref{eq:Annealed-SaddlePointCondition-4}) connecting $\pi$ to $c$:
\begin{equation}\label{eq:mass_action_final_correct}
\ln\!\left( \frac{c (1-\pi)}{\pi (1-c)} \right) = -w_2^{\rm stick} \, q^2 \pi \rho + \frac{w_3^{\rm stick}}{2} \, q^3 \pi^2 \rho^2  \, .
\end{equation}
Notice, in particular, that $\pi$ and $c$ enter in the entropic term of Eq.~\eqref{eq:F_A_final_correct}, which can be viewed then as a form of ``bonding'' entropy generated by the annealed sticker equilibration.
In fact, similar forms of entropy appear in other theories of associating fluids~\cite{Wertheim1984_JStatPhys_I,Wertheim1984_JStatPhys_II,Wertheim1986_JStatPhys_III,JacksonChapmanGubbins1988_MolPhys}.

The phase diagram of the annealed case can then by obtained by a numerical solution of the two equations~\eqref{eq:F_A_final_correct} and~\eqref{eq:mass_action_final_correct}, as it will be illustrated further below (see Sec.~\ref{sec:Results}).
To further rationalize those results, it proves convenient to consider the two opposite limits of weak and strong stickers-stickers interaction.

%%%
(i) {\it Weak sticker-sticker interactions} --
%%%
This condition holds under the assumption that sticker-sticker interactions are weak on the density scale, {\it i.e.}
\begin{equation}\label{eq:weak_assumption}
w_2^{\rm stick} q^2 \rho \ll 1 \, , \qquad w_3^{\rm stick} q^3 \rho^2 \ll 1 \, .
\end{equation}
Then, the probability $\pi$ (Eq.~\eqref{eq:OccBindingProb}) remains close to the bare fraction $c$ with dominant linear corrections in $\rho$.
Using the latter, it is easy to verify that the leading behavior of the mixing term of the free energy density~\eqref{eq:F_A_final_correct} is ${\mathcal O}(\rho^3)$,
\begin{equation}\label{eq:mixing_small_delta_nohalf}
\rho \left[ \pi \ln\!\left(\frac{\pi}c\right) + (1-\pi)\ln\!\left(\frac{1-\pi}{1-c}\right) \right] \simeq \frac{A^2}{2c(1-c)} \, \rho^3 \, ,
\end{equation}
and that, using Eq.~\eqref{eq:mass_action_final_correct}, $A=w_2^{\rm stick} q^2 c^2(1-c)$.
Therefore, up to ``${\mathcal O}(\rho^3)$''-terms, the free energy reads
\begin{eqnarray}\label{eq:FreeEnergyAnnealedWeak}
\frac{\beta F_A}V
& \simeq & \frac{\rho}N \ln\!\left( \frac{\rho}{eN} \right) + \frac12 \left( w_2 - w_2^{\rm stick} q^2 c^2 \right) \rho^2 \nonumber\\
& & + \frac16 \left( w_3 + w_3^{\rm stick} q^3 c^3 - 3 (w_2^{\rm stick})^2 q^4 c^3(1-c) \right) \rho^3 \, . \nonumber\\
\end{eqnarray}
Hence, in the weak-bonding regime, sticker-sticker interactions primarily lead to renormalization of the effective two- and three-body interaction coefficients.
Interestingly, the renormalization of the three-body term receives a contribution from the second-body term of the sticker-sticker interactions (see Appendix~\ref{sec:Relation2SR} for a discussion of a similar effect appearing in a generalized Semenov-Rubinstein theory).

%%%
(ii) {\it Strong sticker-sticker interactions} --
%%%
In the opposite limit
\begin{equation}\label{eq:strong_assumption}
w_2^{\rm stick} q^2 \rho \gg 1 \, , \qquad w_3^{\rm stick} q^3 \rho^2 \gg 1 \, ,
\end{equation}
Eq.~\eqref{eq:mass_action_final_correct} implies that $\pi\to 1$ with exponentially small corrections in $\rho$.
Therefore, one can substitute $\pi=1$ everywhere in Eq.~\eqref{eq:lnZ_hom_annealed} which becomes:
\begin{eqnarray}\label{eq:f_pi_equals_one}
\frac{\beta F_A}V
& \simeq & \frac{\rho}N \ln\!\left( \frac{\rho}{eN} \right) + \frac12 \left( w_2 - w_2^{\rm stick}q^2 \right)\rho^2 \nonumber\\
& & + \frac16 \left( w_3 + w_3^{\rm stick}q^3 \right)\rho^3 + \rho \ln(1/c) \, .
\end{eqnarray}
We notice, then, an analogous renormalization of interaction coefficients as in the case of weak interactions, accompanied by a linear term in $\rho$ equivalent to the coupling of a ``chemical'' potential equal to $-\ln(1/c)$.

%%%
\subsubsection{Quenched model}\label{sec:SPapproxQuenched}
%%%
Similarly to the approach used for the annealed model (Sec.~\ref{sec:SPapproxAnnealed}), we assume here the spatially-homogeneous and replica-independent fields,
\begin{equation}\label{eq:RS_ansatz_corrected}
\rho^{(a)}({\vec r}) = \rho \, , \,\,\,
\phi^{(a)}({\vec r}) = \phi \, , \,\,\,
Q^{(a)}({\vec r}) = Q \, , \,\,\,
\lambda^{(a)}({\vec r}) = \lambda \, ,
\end{equation}
for $a=1, \dots, n$.
Based on that, the replica-partition function Eq.~\eqref{eq:S72} reads:
\begin{equation}\label{eq:ReplicaPartFunct-HomogeneousFields}
\overline{Z^n} = \frac1{(K!)^n} \, \exp\!\Big[ n\left( {\mathcal G}_{\rm vol}(\rho, \phi) + {\mathcal G}_{\rm stick}(Q, \lambda) \right) \Big] \Big[ \zeta_n (\phi, \lambda) \Big]^K \, ,
\end{equation}
with (see Eq.~\eqref{eq:S75}):
\begin{equation}\label{eq:ZetanQuenched-Uniform}
\zeta_n(\phi, \lambda) = V^n \exp\!\left[ -{\rm i} nN \phi + N \ln\!\Big( (1-c) + c \, e^{-{\rm i} q n \lambda} \Big) \right] \, ,
\end{equation}
and with ${\mathcal G}_{\rm vol}(\rho, \phi)$ and ${\mathcal G}_{\rm stick}(Q, \lambda)$ given by the same expressions Eq.~\eqref{eq:G0_hom_here} and~\eqref{eq:GQ_hom_here}, respectively.
Then, using the replica trick~Eq.~\eqref{eq:replica_trick} and the Stirling formula, the free energy of the system per unit volume is given by:
\begin{widetext}
\begin{eqnarray}\label{eq:lnZ_hom_quenched}
\frac{\beta F_{\rm Q}}V 
& \equiv & -\frac1V \lim_{n\to0} \frac{\overline{Z^n} - 1}n \nonumber\\
& = & \frac{K}V\ln\!\left( \frac{K}{eV} \right) - {\rm i} \phi\!\left( \rho-\frac{KN}V \right) + \frac{w_2}2 \rho^2 + \frac{w_3}6 \rho^3 - {\rm i} \lambda\!\left( Q - qc \frac{NK}V \right) - \frac{w_2^{\rm stick}}2 Q^2 + \frac{w_3^{\rm stick}}6 Q^3 \, .
\end{eqnarray}
\end{widetext}

As for the annealed model, we minimize Eq.~\eqref{eq:lnZ_hom_quenched} with respect to $\phi$, $\lambda$ and $Q$, {\it i.e.}
\begin{eqnarray}
\partial_{\phi}(\beta F_{\rm Q} / V) = 0 & \Rightarrow & \frac{K N}V = \rho \, , \label{eq:Quenched-SaddlePointCondition-1} \\
\partial_{\lambda}(\beta F_{\rm Q} / V) = 0 & \Rightarrow & Q = q c \frac{K N}V = q c \rho \, , \label{eq:Quenched-SaddlePointCondition-3} \\
\partial_{Q}(\beta F_{\rm Q} / V) = 0 & \Rightarrow & -{\rm i} \lambda = w_2^{\rm stick}Q - \frac{w_3^{\rm stick}}2 Q^2 \, . \label{eq:Quenched-SaddlePointCondition-4}
\end{eqnarray}
With these conditions, the free energy~\eqref{eq:lnZ_hom_quenched} becomes:
\begin{eqnarray}\label{eq:F_Q_final_correct}
\frac{\beta F_{\rm Q}}V
& = & \frac{\rho}N \ln\!\left( \frac{\rho}{eN} \right) + \frac12 (w_2 - w_2^{\rm stick} q^2 c^2)\rho^2 \nonumber\\
& & + \frac16 (w_3 + w_3^{\rm stick} q^3 c^3) \rho^3 \, .
\end{eqnarray}
In comparison to the expression~\eqref{eq:F_A_final_correct} of the annealed model, Eq.~\eqref{eq:F_Q_final_correct} does not contain any mixing entropy term which remains a genuine feature of the annealed statistics.

%%%
\subsubsection{Comparison of the annealed and quenched models}\label{sec:comparison}
%%%
As a final point, it is interesting to note that Eq.~\eqref{eq:F_Q_final_correct} reduces to Eq.~\eqref{eq:FreeEnergyAnnealedWeak} of the annealed model in the limit of weak sticker-sticker interactions, where the probability $\pi$ for a monomer to be a sticker (Eq.~\eqref{eq:OccBindingProb}) is equal to the bare one, $c$, and where sticker-mediated interactions contribute only through a renormalization of the virial coefficients.
In this respect, the effective two-body virial coefficient $w_2^{\rm eff}$ appearing in the free energy density~\eqref{eq:F_Q_final_correct} changes sign at $w_2^{\rm stick} = w_2 / (q^2 c^2)$ thus switching from dominant repulsive ($w_2^{\rm eff}>0$) to dominant attractive  ($w_2^{\rm eff}<0$) interactions.
Yet, this change alone does {\it not} necessarily imply a thermodynamic phase transition of the polymer solution.
In fact, the actual stability of the homogeneous phase also depends on the translational entropy term and on the higher-order (stabilizing) virial contribution.
A genuine liquid-liquid demixing instability appears only when the homogeneous free energy develops a critical point $(\rho, w_2^{\rm stick}) = (\rho_c, w_{2,c}^{\rm stick})$ and, below it, a spinodal/binodal structure~\cite{RubinsteinColby2003_PolymerPhysics}.
From the chemical potential, $\mu_Q = \partial(F_Q / V) / \partial\rho$, the spinodal critical line as a function of $\rho$ is obtained by imposing $\partial\mu_Q / \partial\rho = 0$, when this condition can be satisfied for the given sticker-sticker interactions.
Then, the critical point is obtained by imposing the further condition, $\partial^2\mu_Q / \partial\rho^2 = 0$, on the second derivative.
Taken these two conditions together, we get the analytical expressions:
\begin{eqnarray}
\rho_c & = & \frac1{\sqrt{N (w_3 + w_3^{\rm stick} q^3 c^3)}} \, , \label{eq:RhoCrit-Quenched} \\
w_{2,c}^{\rm stick} & = & \frac1{q^2 c^2} \left( w_2 + \frac2{N\rho_c} \right) \, . \label{eq:w2StickCrit-Quenched}
\end{eqnarray}
Notice that, since $w_3 + w_3^{\rm stick} q^3 c^3 > 0$, one has that $w_{2,c}^{\rm stick} > w_2 / (q^2 c^2)$, {\it i.e.} the onset of an effective sticker-sticker attraction is not sufficient {\it per se} to produce phase separation.
Then, the spinodal critical line $w_2^{\rm stick}$ as a function of $\rho$ can be written in the following simple form:
\begin{equation}\label{eq:CriticalLine}
\frac{ w_2^{\rm stick} - \frac{w_2}{q^2c^2} }{ w_{2,c}^{\rm stick} - \frac{w_2}{q^2c^2} } = \frac12 ( \rho / \rho_c + (\rho / \rho_c)^{-1} ) \, .
\end{equation}
%

%%%
\section{Phase diagram: numerical methods}\label{sec:PhaseDiagramNumerics}
%%%
In this Section, we describe a simple numerical procedure to determine the critical point and the coexistence curve in the parameters space ``$(\rho, w_2^{\rm stick})$'' for both, the annealed (Eq.~\eqref{eq:F_A_final_correct}) and the quenched model (Eq.~\eqref{eq:F_Q_final_correct}).
The other parameters of the model are maintained fixed and chosen as it follows.
Taking $b$ as our length unit, we fix $w_2 / b^3 = 1$ and $w_3 / b^6 = w_3^{\rm stick} / b^6 = 1$ throughout the rest of the paper.
The dimensionless sticker charge is also fixed, $q=1$.
Finally, we consider sticker fractions $c=0.25,0.50,0.75$.

%%%
\subsection{Annealed model}\label{sec:PhaseDiagramNumerics-Annealed}
%%%

%
\begin{figure}
\includegraphics[width=0.45\textwidth]{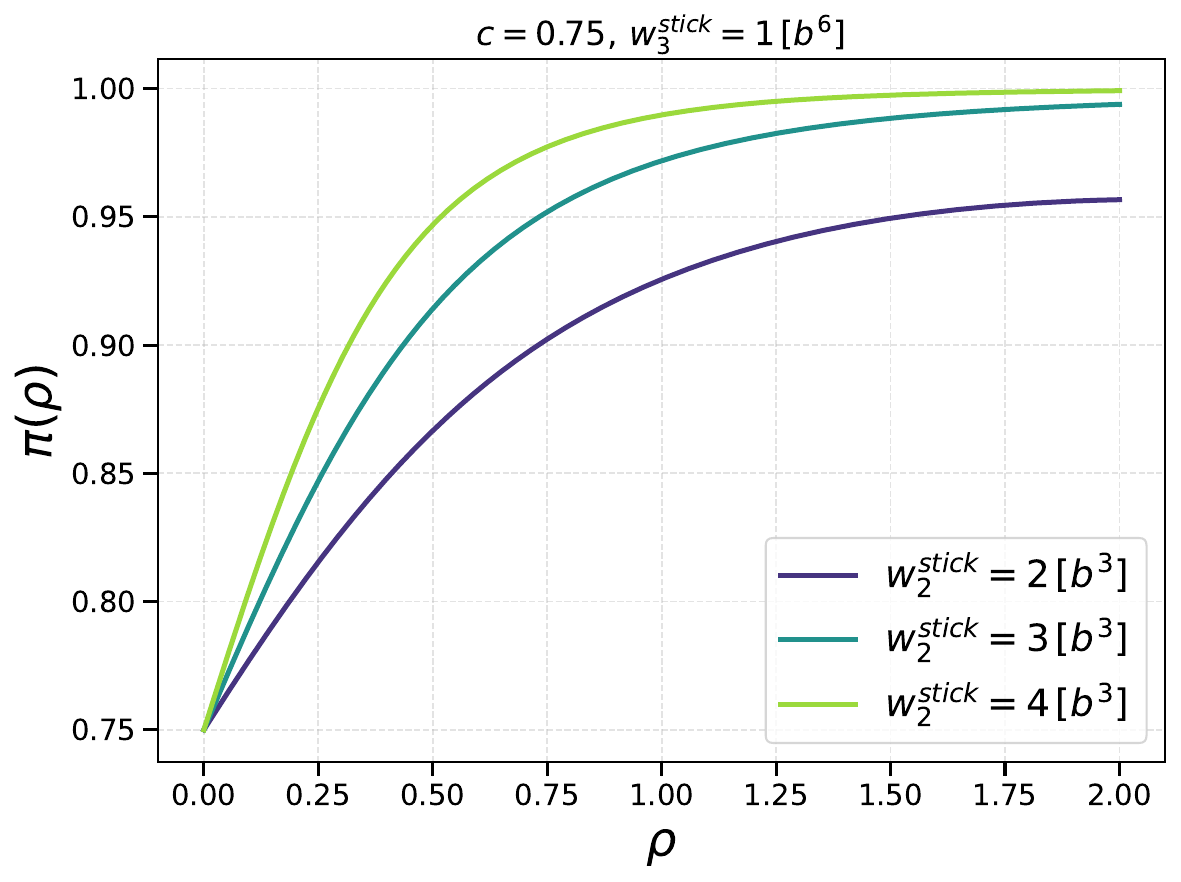}
\caption{
Sticker probability $\pi$ (Eq.~\eqref{eq:mass_action_final_correct}) as a function of monomer density $\rho$, for $w_2 / b^3 = w_3 / b^6 = w_3^{\rm stick} / b^6 = 1$, $q = 1$ and for $c=0.75$ and $w_2^{\rm stick} / b^3 = 2,3,4$ (see legend).
The probability $\pi$ is bounded as $c \leq \pi \leq 1$, and for the same $\rho$ it increases as a function of the sticker attraction strength $w_2^{\rm stick}$.
}
\label{fig:pirho}
\end{figure}

For the annealed case, the probability $\pi = \pi(w_2^{\rm stick}, \rho)$ is first computed by solving the mass-action equation~\eqref{eq:mass_action_final_correct} at fixed control parameter $w_2^{\rm stick}$ and density $\rho$ (see Fig.~\ref{fig:pirho}, showing $\pi$ as a function of $\rho$, for $c=0.75$ and for $w_2^{\rm stick}/b^3 =2,3,4$).
Then, the solution is inserted into the annealed free energy density $\beta F_A / V$~\eqref{eq:F_A_final_correct}, from which the chemical potential,
\begin{equation}\label{eq:ChemicalPotential}
\mu_A(\rho) \equiv \frac1V \frac{\partial F_A}{\partial\rho} \, ,
\end{equation}
is evaluated.
The critical point is located from the merging of the two spinodal branches, identified as the zeros of $\partial\mu_A / \partial\rho$.
The coexistence curve is subsequently constructed by imposing the common-tangent condition on the free energy density and solving the resulting nonlinear system for the two coexistence densities. 
As known, this procedure is thermodynamically equivalent to requiring equality for both the chemical potential $\mu_{A}(\rho)$ and the osmotic pressure,
\begin{equation}\label{eq:OsmoticPressure}
\Pi_A(\rho) \equiv \rho \, \mu_A(\rho) - F_A / V \, ,
\end{equation}
in the two coexisting phases, {\it i.e.}
\begin{equation}\label{eq:PD}
\mu_A(\rho_1) = \mu_A(\rho_2), \qquad \Pi_A(\rho_1) = \Pi_A(\rho_2) \, .
\end{equation}
\begin{figure*}
\includegraphics[width=0.85\textwidth]{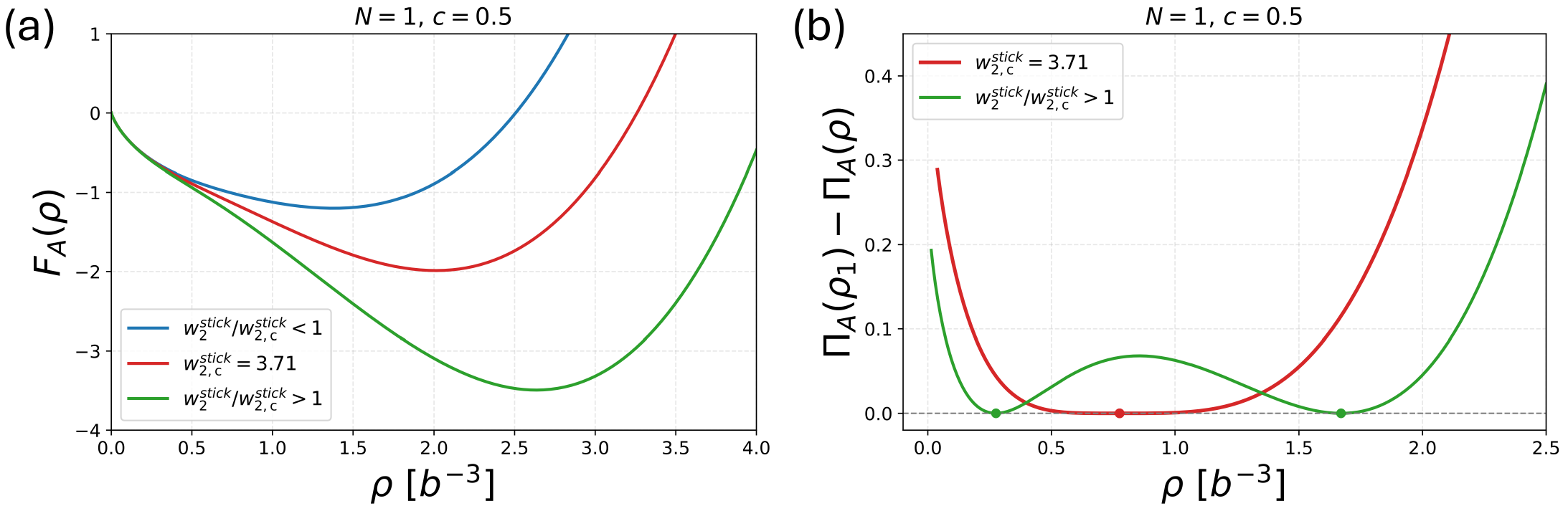}
\caption{
(a)
Free energy profiles for the annealed model $F_A(\rho)$ (Eq.~\eqref{eq:F_A_final_correct} with the solution of the mass-action relation Eq.~\eqref{eq:mass_action_final_correct}) as a function of the monomer density $\rho$, for $N=1$ and $c=0.5$ and parameters $w_2 / b^3 = 1$, $w_3 / b^6 = w_3^{\rm stick} / b^6 = 1$ and $q=1$ (see Sec.~\ref{sec:PhaseDiagramNumerics} for details). 
The profiles are for representative values of the sticker-mediated attraction strength $w_2^{\rm stick}$ smaller than (blue line), equal to (red line), and larger than (green line) the critical value $w_{2,c}^{\rm stick} \simeq 3.71$ (see legend). 
(b)
Corresponding osmotic pressure profiles $\Pi_A(\rho_1) - \Pi_A(\rho)$ (Eq.~\eqref{eq:OsmoticPressure}) as a function of the monomer density $\rho$, measured at the coexistence of the two stable phases $\rho_1 < \rho_2$.
The dashed horizontal line marks the coexistence condition $\Pi_A(\rho_1) = \Pi_A(\rho_2)$, and the filled circles indicate the positions of the minima.
}
\label{fig:Comparing_FreeEnergies+GrandPotentials}
\end{figure*}

Fig.~\ref{fig:Comparing_FreeEnergies+GrandPotentials} shows an example of this geometric construction for the annealed system with $N=1$ and $c=0.5$ and the rest of the parameters fixed as reported in the beginning of this Section.
After having determined the critical value $w_2^{\rm stick} = w_{2,c}^{\rm stick} \simeq 3.71$, the profiles for $\beta F_A(\rho) / V$ (panel (a)) show two inflexion points only for $w_2^{\rm stick} / w_{2,c}^{\rm stick} > 1$ as expected.
Accordingly (panel (b)), the osmotic pressure profiles $\Pi_A(\rho_1) - \Pi_A(\rho)$ for $\mu_A$ taken at the coexistence value and with $\rho_1$ the low-density minimum as a reference show one minimum at the critical point (red line) and two minima for $w_2^{\rm stick} / w_{2,c}^{\rm stick} > 1$ (green lines).

%%%
\subsection{Quenched model}\label{sec:PhaseDiagramNumerics-Quenched}
%%%
For the quenched case, the sticker-distribution-average free energy density $\beta F_Q / V$ (Eq.~\eqref{eq:F_Q_final_correct}) depends only on $\rho$ through effective renormalized coefficients.
In this case, the critical point was obtained analytically (see Eqs.~\eqref{eq:RhoCrit-Quenched} and~\eqref{eq:w2StickCrit-Quenched}) and the binodal line is again determined through the common-tangent construction as described in Sec.~\ref{sec:PhaseDiagramNumerics-Annealed}.

%%%
\section{Results}\label{sec:Results}
%%%

%
\begin{figure*}
\includegraphics[width=0.75\textwidth]{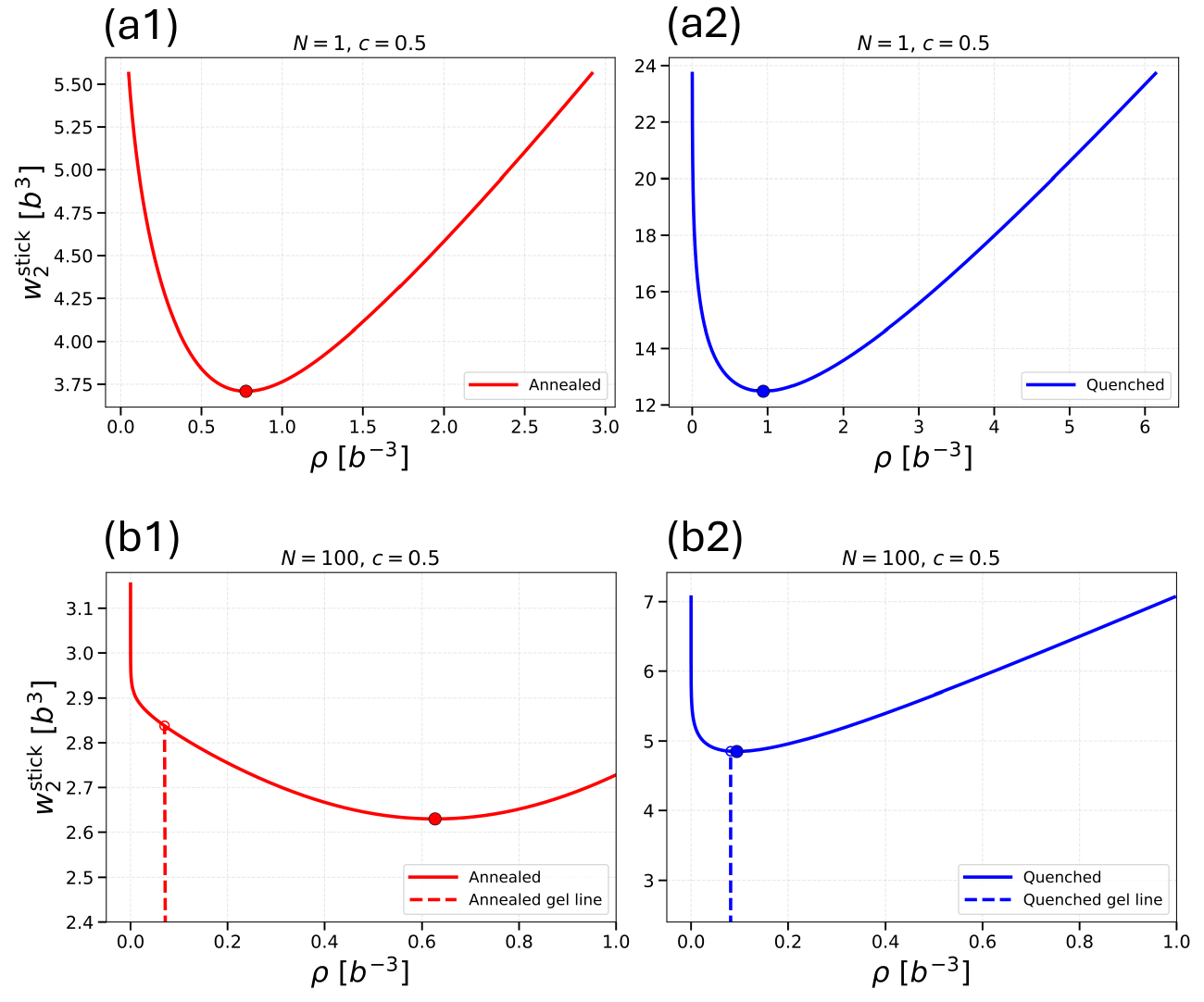}
\caption{
Annealed (red lines) and quenched (blue lines) phase diagrams in the ``monomer density / sticker-mediated attraction strength $(\rho, w_2^{\rm stick})$'' plane, for sticker fraction $c = 0.50$ and parameters 
$w_2/b^3 = 1$, $w_3 / b^6 = w_3^{\rm stick} / b^6 = 1$ and $q=1$ (see Sec.~\ref{sec:PhaseDiagramNumerics} for details).
(a1, a2) Results for polymer chains of $N = 1$ monomers.
(b1, b2) Results for polymer chains of $N = 100$ monomers.
In each panel, the position of the critical point is marked by the filled circle.
For $N=100$, the dashed lines denote the sol-gel transition (from left to right). 
Increasing $N$ suppresses the translational entropy contribution and favors phase separation.
}
\label{fig:PhaseDiagrams_RescaledNo}
\end{figure*}

Fig.~\ref{fig:PhaseDiagrams_RescaledNo} shows curves of phase diagrams in the ``monomer density / sticker-mediated attraction strength $(\rho, w_2^{\rm stick})$'' plane, for associating polymers with annealed (red solid lines) and quenched (blue solid lines) sticker disorder and sticker fraction $c = 0.50$.
Panels (a1) and (a2) are for the limiting case of very short polymers ($N=1$), while panels (b1) and (b2) are for long polymers ($N=100$). 

In all considered cases the system undergoes a liquid-liquid phase separation bounded by a binodal terminating at a critical point (filled circle), arising from the competition between excluded-volume repulsion and sticker-mediated interactions.
The parameter $w_2^{\rm stick}$ controls the effective bonding strength and therefore plays a role analogous to an inverse temperature: increasing $w_2^{\rm stick}$ enhances the attractive contribution associated with sticker-mediated interactions and thus promotes phase separation.
In the annealed theory the variable $\pi$, which can be interpreted as a {\it dressed} probability for a monomer to behave as a sticker (see Eq.~\eqref{eq:mass_action_final_correct}), is not fixed a priori but is determined self-consistently through the local thermodynamic balance (see Fig.~\ref{fig:pirho}). 
As a consequence, the annealed free energy~\eqref{eq:F_A_final_correct} contains an additional $\pi$-dependent mixing-entropy contribution with respect to the quenched free energy~\eqref{eq:F_Q_final_correct}, which reflects the configurational freedom associated with sticker equilibration and stabilizes the homogeneous mixed phase.
Physically, the system can optimize the fraction of active stickers so as to reduce the free energy cost of mixing.
As a result, weaker attraction is required to induce demixing than in the quenched case where the sticker fraction $c$ is fixed.
Consequently, the annealed binodal curve lies always {\it below} the corresponding quenched one, indicating that phase separation occurs at weaker attraction strength in the annealed system, in line with an intuitive picture (note the different scales in the horizontal axis of Fig.~\ref{fig:PhaseDiagrams_RescaledNo}). 

Furthermore, in both the annealed and quenched models the dependence on $N$ enters through the chain-entropic contribution, $\frac{\rho}N \ln( \frac{\rho}{eN} )$, whereas the interaction terms retain the same functional form in $\rho$.
By varying $N$, the overall topology of the phase diagram remains unchanged, with a single binodal terminating at a critical point, but instead it changes the relative balance between entropy and attractive interactions.
For large $N$, the translational/chain entropy per monomer is significantly reduced, so that the entropic penalty opposing phase separation becomes weaker.
The net result is that phase coexistence is stabilized already at lower values of the attractive control parameter.

Finally, following~\cite{SemenovRubinstein1998_Statics,RubinsteinColby2003_PolymerPhysics} we have computed the {\it sol-gel} transition line using the Flory criterion $p = 1/(f-1)$ where $p$ is the fraction of stickers which are associated in pairs and $f$ is the number of stickers on the single polymer chain.
Taking (i) $p=b^3\rho \, \pi^2$ and $f=\pi N$ for the annealed system and (ii) $p=b^3\rho \, c^2$ and $f=cN$ for the quenched system, the corresponding sol-gel lines are easily obtained.
The results are shown in panels (b1) and (b2) of Fig.~\ref{fig:PhaseDiagrams_RescaledNo} (dashed lines).
In particular, as one would intuitively expect, no sol-gel transition is possible for $N=1$.

\begin{figure*}
\includegraphics[width=0.75\textwidth]{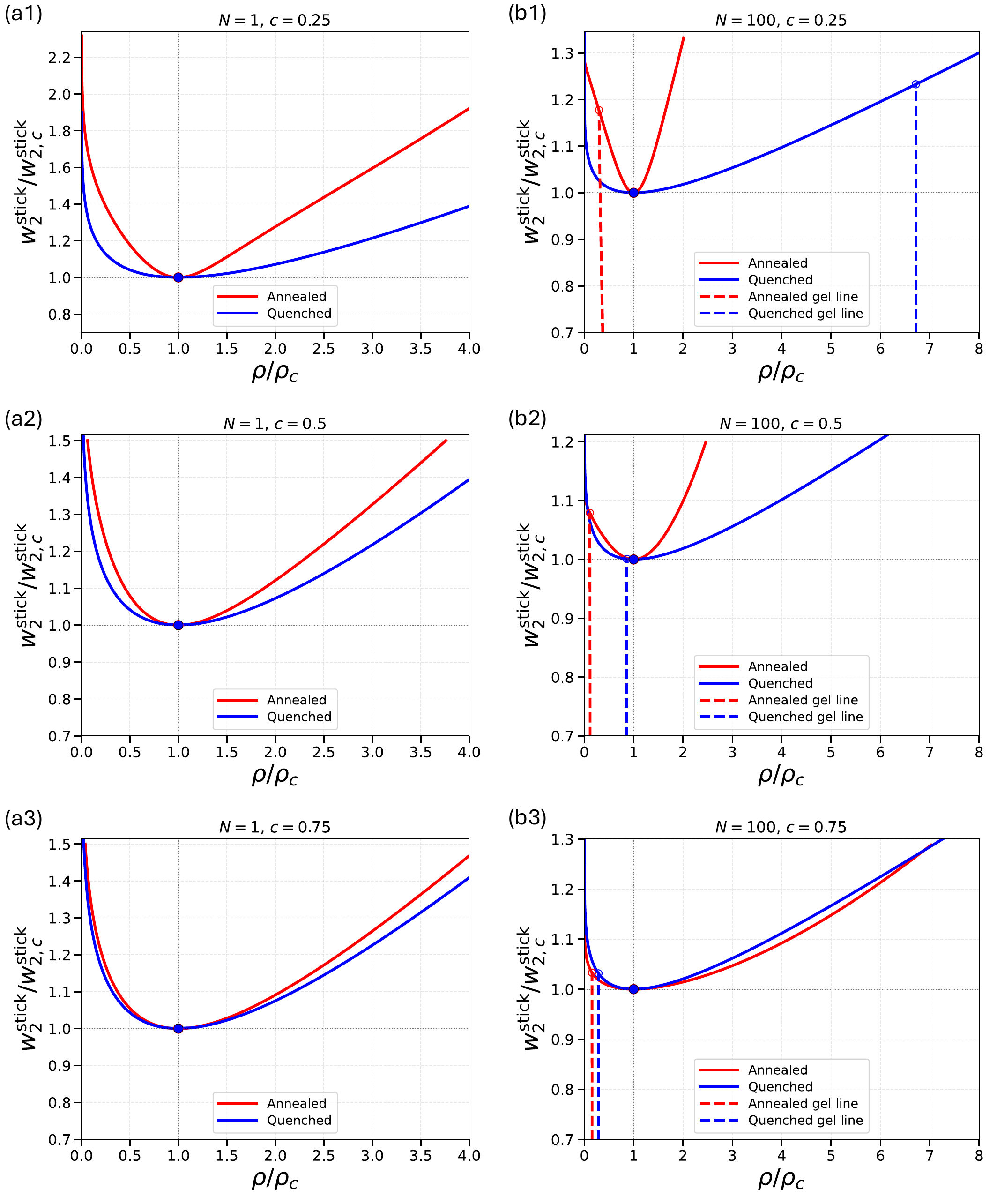}
\caption{
Annealed (red lines) and quenched (blue lines) phase diagrams in the ``monomer density / sticker-mediated attraction strength $(\rho, w_2^{\rm stick})$'' plane and with $x$- and $y$-axes given in units of critical values $\rho_c$ and $w_{2,c}^{\rm stick}$ (the position of the critical point $(1,1)$ is indicated by the filled circle).
We consider sticker fractions $c = 0.25, 0.50, 0.75$ and parameters 
$w_2/b^3 = 1$, $w_3 / b^6 = w_3^{\rm stick} / b^6 = 1$ and $q=1$ (see Sec.~\ref{sec:PhaseDiagramNumerics} for details).
(a1-a3) Results for polymer chains of $N = 1$ monomers.
(b1-b3) Results for polymer chains of $N = 100$ monomers.
For $N=100$, the dashed lines mark the sol-gel transition (from left to right).
}
\label{fig:PhaseDiagrams_RescaledYes}
\end{figure*}

We have extended this analysis to lower ($c = 0.25$) and higher ($c = 0.75$) values of sticker fractions.
The results are shown in Fig.~\ref{fig:PhaseDiagrams_RescaledYes} where, to ease comparison, we employ the representation ``$(\rho / \rho_c, w_2^{\rm stick} / w_{2,c}^{\rm stick})$'' in normalized variables with respect to critical values.
Two important observations stand out.
First, in the limit $c\to 1$ disorder is effectively suppressed, annealed and quenched statistics become equivalent, so phase boundaries overlap (panels (a3) and (b3)).
Second (and, more importantly) by increasing the sticker fraction $c$ in the quenched model, the critical point moves from the sol-region into the gel-region (namely from the left to the right of the dashed blue line, see panels from (b1) to (b3)).
This is a testable prediction which could be easily verified in, {\it e.g.}, computer simulations of associating polymers~\cite{Chen2025sol, RovigattiSciortino2022}.

%%%
\section{Discussion and conclusions}\label{sec:DiscConcls}
%%%
In this work, we have formulated a unified microscopic density-functional theory for associating polymers with reversible sticker interactions.
The theory places annealed and quenched disorder on the same microscopic footing while revealing their fundamentally different thermodynamic consequences.
Annealed disorder (free energy Eq.~\eqref{eq:F_A_final_correct}) gives rise to an additional collective binding field ($\pi$, Eq.~\eqref{eq:OccBindingProb}) that acts as an emergent thermodynamic variable, whereas quenched disorder (free energy Eq.~\eqref{eq:F_Q_final_correct}) enters primarily through a renormalization of excluded-volume interactions.
Importantly, the formulation retains a direct connection to the underlying interacting-chain Hamiltonian (see Sec.~\ref{sec:PolymerModel-Definition}), with approximations introduced only at clearly identifiable stages of the field-theoretic analysis.

An important advantage of the present field-theoretic formulation is that it provides a controlled starting point for incorporating fluctuation corrections beyond mean-field.
This is particularly relevant in light of growing evidence that associating polymers and biomolecular condensates may exhibit critical-like behavior near phase boundaries, including enhanced fluctuations, broad droplet-size distributions, and increasing correlation lengths~\cite{Amico2024,VendruscoloFuxreiter2023}.
Extending the theory beyond the saddle-point approximation would therefore enable a quantitative characterization of correlation lengths and critical fluctuations, providing a framework to connect microscopic sticker statistics with the emergence of discontinuous versus critical phase-separation behavior.

An experimentally relevant consequence of our theory is the qualitatively different role played by annealed and quenched disorder in determining phase behavior.
In particular, the two realizations lead to distinct coexistence boundaries and critical parameters, while only the annealed case supports a self-consistently determined bonding order parameter associated with the equilibration of sticker identities.

Our theoretical phase diagrams can be tested in coarse-grained molecular dynamics simulations, by implementing explicit reversible sticker-sticker potentials, as in recent sticker-spacer models of associative biomolecular polymers and related designs of specific entropic binding~\cite{Chen2025sol, RovigattiSciortino2022}.
At the same time, the predictions suggest concrete routes for experimental tests in systems where reversible interactions can be programmed with a high degree of control.
In particular, DNA-mediated associating materials provide a promising setting in this regard.
Limited-valence DNA nanostars offer direct access to phase boundaries and critical phenomena~\cite{Biffi2013_PNAS_DNAnanostars}, while recent advances in programmable DNA interactions enable the realization of both fixed and dynamically reconfigurable bonding architectures~\cite{Karmakar2025}. 
Such systems may therefore serve as model platforms for probing the thermodynamic consequences of disorder statistics in associating polymer solutions and for assessing the distinct signatures predicted here for annealed and quenched disorder.

In summary, our approach opens the way to quantitative studies of complex reversible networks, gelation phenomena, and spatially heterogeneous associating fluids within a fully field-theoretical framework.

%%%
%\acknowledgements
%%%

%%%
\appendix
%%%

%%%
\section{Field-theoretic formulation of polymer's propagator and saddle point approximation}\label{sec:feynman}
%%%
Eqs.~\eqref{eq:zeta_phi_lambda_cont} and~\eqref{eq:S75} are equivalent to the partition function of a Gaussian chain or $n$ Gaussian chains in an external, time-independent field $\mathcal V$.
Under these conditions, it is known~\cite{DoiEdwards} that the probability amplitude that a single chain of $N$ monomers with the first monomer of index ``$0$'' at spatial position $\vec r_0$ and last monomer of index ``$N$'' at spatial position $\vec r_N$ has the form of the propagator
\begin{equation}\label{eq:DefinePropagator}
G_{\mathcal V}(\vec r_N, N | \vec r_0, 0) = \langle \vec r_N | e^{-N {\mathcal H}} | \vec r_0 \rangle \, ,
\end{equation}
with ``Hamiltonian'' ${\mathcal H} = -\frac{b^2}6\nabla^2 + {\mathcal V}$.

Now consider the Laplace transform,
\begin{eqnarray}\label{eq:Gomega}
\widetilde G_{\mathcal V}(\vec r_N, \vec r_0; \omega)
& = & \int_0^{\infty} {\rm d}N \, G_{\mathcal V}(\vec r_N, N | \vec r_0, 0) \, e^{-\omega N} \nonumber\\
& = & \langle \vec r_N | (\omega\mathbb{I} + {\mathcal H})^{-1} | \vec r_0 \rangle \, ,
\end{eqnarray}
and let $\{ \varphi_{\alpha=1,\dots,n} \} \equiv (\varphi_1(\vec r), \dots, \varphi_n(\vec r))$ be a real $n$-component field\footnote{``$n$'' here ought not be confused with the ``$n$'' in Sec.~\ref{sec:quenched_continuum} denoting the number of replicas.} and
\begin{equation}\label{eq:DefineAction}
A_n(\{ \varphi_{\alpha} \}) = \frac12 \int {\rm d}\vec r \, \sum_{\alpha=1}^n \varphi_\alpha(\vec r) \left( \omega\mathbb{I} + {\mathcal H} \right) \varphi_\alpha(\vec r) 
\end{equation}
be the corresponding action with $\mathbb{I}$ the identity operator.
From standard properties of Gaussian integrals, we know that
\begin{align}\label{eq:<phiXphiX0>}
& \left\langle \varphi_\alpha(\vec r_N) \varphi_\alpha(\vec r_0) \right\rangle \nonumber\\
& \equiv \frac1{Z(\omega)^n} \int \mathcal D\varphi_1 \dots \mathcal D\varphi_n \, e^{-A_n(\{\varphi_\alpha\})} \varphi_\alpha(\vec r_N) \varphi_\alpha(\vec r_0) \nonumber\\
& = \langle \vec r_N | ( \omega\mathbb{I} + {\mathcal H} )^{-1} | \vec r_0 \rangle \, ,
\end{align}
where
\begin{eqnarray}\label{eq:Zomega}
Z(\omega)
& = & \int \mathcal D\varphi \, e^{-\frac12 \int {\rm d}\vec r \, \varphi(\vec r) \left( \omega\mathbb{I} + {\mathcal H} \right) \varphi(\vec r)} \nonumber\\
& = & (\det(\omega\mathbb{I} + {\mathcal H}))^{-1/2} \, .
\end{eqnarray}
Since Eq.~\eqref{eq:<phiXphiX0>} is valid for any $n$ (and $\alpha$), in the limit $n\to0$ we obtain (see Eq.~\eqref{eq:Gomega})
\begin{align}
& \widetilde G_{\mathcal V}(\vec r_N, \vec r_0; \omega) = \langle \vec r_N | ( \omega\mathbb{I} + {\mathcal H} )^{-1} | \vec r_0 \rangle \nonumber\\
& = \lim_{n \to 0} \int \mathcal D\varphi_1 \cdots \mathcal D\varphi_n \, e^{-A_n(\{\varphi_\alpha\})} \varphi_1(\vec r_N) \varphi_1(\vec r_0) \, ,
\end{align}
which gets rid of the partition function $Z(\omega)$.
Thanks to this, the inverse Laplace transform is easily performed:
\begin{equation}
G_{\mathcal V}(\vec r_N, N | \vec r_0, 0) = \int_{\gamma} \frac{{\rm d}\omega}{2\pi {\rm i}} \, \widetilde G_{\mathcal V}(\vec r_N, \vec r_0; \omega) \, e^{\omega N} \, ,
\end{equation}
where $\gamma$ is a straight contour in the complex plane parallel to the imaginary axis in the region of convergence of $\widetilde G_{\mathcal V}(\vec r_N, \vec r_0; \omega)$.
Using Eq.~\eqref{eq:DefineAction}, this leads to
\begin{widetext}
\begin{eqnarray}
G_{\mathcal V}(\vec r_N, N | \vec r_0,0)
& = & \lim_{n \to 0} \int \mathcal D\varphi_1 \cdots D\varphi_n \, e^{-\frac12 \int {\rm d}\vec r \sum_\alpha \varphi_\alpha(\vec r) {\mathcal H} \varphi_\alpha(\vec r)} \, \varphi_1(\vec r_N) \varphi_1(\vec r_0) \int_{\gamma} \frac{d\omega}{2\pi {\rm i}} \, e^{\omega \left( N - \frac12 \int{\rm d}\vec r \, \sum_\alpha \varphi_{\alpha}(\vec r)^2 \right)} \nonumber\\
& = & \lim_{n \to 0} \int \mathcal D\varphi_1 \cdots D\varphi_n \, e^{-\frac12 \int {\rm d}\vec r \sum_\alpha \varphi_\alpha(\vec r) {\mathcal H} \varphi_\alpha(\vec r)} \, \varphi_1(\vec r_N) \varphi_1(\vec r_0)  \, \delta\!\left( N - \frac12 \int {\rm d}\vec r \, \sum_\alpha \varphi_{\alpha}(\vec r)^2 \right) \, .
\end{eqnarray}
\end{widetext}
Redefining $\varphi_\alpha \to \sqrt{N} \, \varphi_\alpha$ and taking into account that $\lim_{n\to0} N^{n/2} \to 1$, one finally obtains
\begin{widetext}
\begin{eqnarray}\label{eq:delta}
G_{\mathcal V}(\vec r_N, N | \vec r_0, 0)
& = &
\lim_{n\to0} \int \mathcal D\varphi_1 \cdots D\varphi_n \, \varphi_1(\vec r_N) \varphi_1(\vec r_0) \nonumber\\
& & \times \exp\!\left[ -\frac{N}2 \int {\rm d}\vec r \, \sum_\alpha \left( \frac{b^2}6(\nabla \varphi_\alpha(\vec r))^2 + {\mathcal V}(\vec r)\varphi_\alpha(\vec r)^2 \right) \right]
\delta\!\left( \frac12\int {\rm d}{\vec r} \, \sum_\alpha \varphi_\alpha(\vec r)^2 - 1 \right) \, .
\end{eqnarray}
\end{widetext}
In the ``large-$N$'' limit, the saddle point approximation leads to the ground-state dominance approximation for $G_{\mathcal V}$, assuming that one minimizes
$$
\int {\rm d}\vec r \, \sum_{\alpha} \left( \frac{b^2}6(\nabla \varphi_\alpha(\vec r))^2 + {\mathcal V}(\vec r) \varphi_\alpha(\vec r)^2 \right) \, ,
$$
with $\varphi_\alpha(\vec r) = \delta_{\alpha,1} \phi(\vec r)$ and $\frac12 \int {\rm d}\vec r \, \phi(\vec r)^2 = 1$, in order to satisfy the constraint.
This is due to the presence of the term ``$\varphi_1(\vec r_N) \varphi_1(\vec r_0)$'' in Eq.~\eqref{eq:delta}, which breaks the $O(n)$-symmetry of the integrand but does not contribute at leading order to the saddle point.
Importantly, the representation~\eqref{eq:delta} allows one to compute corrections to the strict ``large-$N$'' limit.

A particularly simple, yet instructive, example (which is used in Sec.~\ref{sec:SPapproxAnnealedQuenched}) assumes ${\mathcal V}$ as a constant.
Then, $G_{\mathcal V}(\vec r_N, N | \vec r_0, 0)$ depends only on $\vec r \equiv \vec r_N - \vec r_0$ and (see Eq.~\eqref{eq:DefinePropagator})
\begin{eqnarray}\label{eq:time}
\int {\rm d}\vec r_N \, {\rm d}\vec r_0 \, G_{\mathcal V}(\vec r_N, N | \vec r_0, 0)
& = & V \, e^{-N{\mathcal V}} \! \int {\rm d}\vec r \, G_{{\mathcal V}=0}(\vec r, N | \vec 0, 0) \nonumber\\
& = & V \, e^{-N{\mathcal V}} \, ,
\end{eqnarray}
where $V$ is the volume of the system and $G_{{\mathcal V} = 0}(\vec r_N, N | \vec r_0, 0) = \left( \frac3{2\pi b^2 N} \right)^{\!3/2} \exp\!\left[ -\,\frac{3(\vec r_N - \vec r_0)^2}{2b^2N} \right]$ is the ``free'' propagator of the Gaussian chain~\cite{DoiEdwards}.

%%%
\section{Relation to the classical gelation theory by Semenov and Rubinstein}\label{sec:Relation2SR}
%%%
Interestingly, in the weak-bonding regime, our annealed theory (see Eq.~\eqref{eq:FreeEnergyAnnealedWeak}) coincides with a generalization of the classical gelation theory by Semenov and Rubinstein for thermoreversible polymer gels~\cite{SemenovRubinstein1998_Statics}.
To show this, we recall the purely combinatorial derivation of the partition function of an ideal gas of stickers in which
(i) stickers bond reversibly to each other in pairs and, at the same time,
(ii) the addition of a third sticker to an already formed pair (a feature not considered originally by Semenov and Rubinstein) is allowed.
Here, we concentrate only on the contribution to the free energy due to the stickers.
The other contributions, essentially the polymeric part of the free energy (chain translational entropy and ordinary virial terms in the monomer density), remain unchanged with respect to the original Semenov-Rubinstein treatment and are therefore omitted here.

%%%%
%\subsection{Microscopic counting variables}\label{subsec:counting_variables}
%%%%
{\it The model} --
Consider $N_{\rm st}$ stickers in a volume $V$, and assume that stickers can only bound in pairs or triplets.
For a given number of pairs ($N_p$) and triplets ($N_t$) the following relations hold:
\begin{eqnarray}
& & 0 \le N_p \le \lfloor N_{\rm st}/2 \rfloor \, , \label{eq:DefiningIntrusions-1} \\
& & 0 \le N_t \le \min( N_p, N_{\rm st} - 2N_p ) \, , \label{eq:DefiningIntrusions-2}
\end{eqnarray}
where $\lfloor \cdot \rfloor$ denotes the integer part.
Following Semenov and Rubinstein~\cite{SemenovRubinstein1998_Statics}, the total number of configurations with $N_p$ pairs is given by the combinatorial factor:
\begin{equation}\label{eq:P_pair_correct}
\binom{N_{\rm st}}{2N_p}(2N_p-1)!! = \frac{ N_{\rm st}! }{ 2^{N_p} \, N_p! \, (N_{\rm st} - 2N_p)! } \, ,
\end{equation}
where the first term accounts for the choice of the $2N_p$ stickers involved in bonds, while the second is for the number of possible pairings among them.
Then, assuming that one (and, no more than one) amongst the $N_{\rm st}-2N_p$ unpaired stickers may bond to a formed pair to get a triplet, the total number of configurations associated to $N_t$ of these events is given by the following factor:
\begin{align}\label{eq:P_intr_corrected}
& \binom{N_{\rm st} - 2N_p}{N_t} \, \frac{N_p!}{(N_p-N_t)!} \nonumber\\
& = \frac{N_p! \, (N_{\rm st} - 2N_p)!}{N_t! \, (N_p-N_t)! \, (N_{\rm st} - 2N_p - N_t)!} \, ,
\end{align}
where the first term comes from choosing $N_t$ stickers (with the constraint~\eqref{eq:DefiningIntrusions-2} on their available number) from the $N_{\rm st} - 2N_p$ unpaired ones, with the second term constraining each of them to be assigned to one and only one of the $N_p$ already existing pairs.
Multiplying Eq.~\eqref{eq:P_pair_correct} by Eq.~\eqref{eq:P_intr_corrected} and assuming that the Boltzmann weight for forming a pair is $w_p = \frac{v_b}V e^{\beta \epsilon_p}$ and that for a triplet is $w_t = \frac{v_b}V e^{\beta \epsilon_t}$ where $v_b$ is the ``bonding volume'' and $\epsilon_p, \, \epsilon_t$ are energy costs associated to the processes, after some simple manipulations the partition function of the system is finally given by:
\begin{widetext}
\begin{equation}\label{eq:Zst_lambda_theta_corrected}
Z_{\rm st} = \sum_{N_p=0}^{\lfloor N_{\rm st}/2\rfloor} \; \sum_{N_t=0}^{\min(N_p,\;N_{\rm st}-2N_p)} \frac{ N_{\rm st}! }{ 2^{N_p} \, N_t! \, (N_p-N_t)! \, (N_{\rm st} - 2N_p-N_t)! } \, w_p^{N_p} \, w_t^{N_t} \, .
\end{equation}
\end{widetext}
%

%%%%
%\subsection{Large-\(N_{\rm st}\) free energy density}\label{subsec:stirling_phi_correct}
%%%%
{\it Asymptotic behavior} --
The partition function~\eqref{eq:Zst_lambda_theta_corrected} is exact, but difficult to handle.
To simplify the formalism, we study $Z_{\rm st}$ in the thermodynamic limit $N_{\rm st}\to\infty$ and $V\to\infty$ with sticker number density, $\rho_{\rm st} \equiv N_{\rm st} / V$, fixed.
Then, introducing the fractions (corresponding to Eqs.~\eqref{eq:DefiningIntrusions-1} and~\eqref{eq:DefiningIntrusions-2})
\begin{eqnarray}
p & \equiv & \frac{2N_p}{N_{\rm st}} \in [0,1] \, , \label{eq:def_p_t_correct-1} \\
t & \equiv & \frac{N_t}{N_{\rm st}} \, , \,\,\, 0 \le t\le \min \! \left( \frac{p}2, \, 1-p \right) \, , \label{eq:def_p_t_correct-2}
\end{eqnarray}
and using $N_p / V = \frac{p}2 \rho_{\rm st}$ and $N_t / V = t \rho_{\rm st}$ and once again the Stirling approximation $n! \approx (n/e)^n$, we get the following expression for the free energy density:
\begin{equation}\label{eq:Fst_def_correct}
\frac{\beta F_{\rm st}}V = -\frac{\ln ( Z_{\rm st} )}V \approx \rho_{\rm st} \, \inf_{p,t} \phi(p, t; \epsilon_p, \epsilon_t) \, ,
\end{equation}
where
\begin{align}\label{eq:phi_pre_saddle_correct}
\phi(p, t; \epsilon_p, \epsilon_t) & = t \ln(t) + (1-p-t) \ln(1-p-t) \nonumber\\
& + \left( \frac{p}2 - t \right) \ln\!\left( \frac{p}2 - t \right) - \left( \frac{p}2 + t \right) \ln\left( \frac{\rho_{\rm st}v_b}e \right) \nonumber\\
& + \frac{p}2 \ln(2) - \frac{p}2\beta\epsilon_p - t\beta\epsilon_t \, .
\end{align}
Notice that Eq.~\eqref{eq:phi_pre_saddle_correct} is identical to Eq.~(2.11) of Semenov and Rubinstein~\cite{SemenovRubinstein1998_Statics} for $t=0$, that is when the number $N_t$ of triplets vanishes.
This was to be expected.

Differentiating Eq.~\eqref{eq:phi_pre_saddle_correct} with respect to $p$ and $t$ leads to the following equations for the optimal $(p^\ast, t^\ast)$:
\begin{eqnarray}
\rho_{\rm st} v_b \, e^{\beta \epsilon_p} & = & \frac{p^\ast-2t^\ast}{(1-p^\ast-t^\ast)^2} \, , \label{eq:saddle_p_correct} \\
\rho_{\rm st} v_b \, e^{\beta \epsilon_t} & = & \frac{2t^\ast}{(1-p^\ast-t^\ast)(p^\ast-2t^\ast)} \, . \label{eq:saddle_t_correct}
\end{eqnarray}
Then, after substitution of~\eqref{eq:saddle_p_correct} and~\eqref{eq:saddle_t_correct} in~\eqref{eq:phi_pre_saddle_correct}, one gets the following expression for the free energy density:
\begin{equation}\label{eq:Fst_post_saddle_correct}
\frac{\beta F_{\rm st}}V = \rho_{\rm st} \left[ \ln(1-p^\ast-t^\ast) + \frac{p^\ast}2 + t^\ast \right] \, .
\end{equation}
%

%%%%
%\subsection{Weak-bonding / low-density expansion}\label{subsec:weak_bonding_expansion_correct}
%%%%
{\it Low sticker density limit} --
In the regime where stickers are dilute ($\rho_{\rm st} \ll 1$), Eqs.~\eqref{eq:saddle_p_correct} and~\eqref{eq:saddle_t_correct} can be expanded to second-order in $\rho_{\rm st}$, {\it i.e.}
\begin{eqnarray}\label{eq:small_lambda_theta}
p^\ast & \approx & v_b \, e^{\beta\epsilon_p} \rho_{\rm st} + v_b^2 \, e^{2\beta\epsilon_p} (e^{\beta(\epsilon_t-\epsilon_p)} - 2) \, \rho_{\rm st}^2 \, , \label{eq:small_pstar} \\
t^\ast & \approx & \frac12 \, v_b^2 \, e^{\beta(\epsilon_p+\epsilon_t)} \, \rho_{\rm st}^2 \, . \label{eq:small_tstar} 
\end{eqnarray}
Correspondingly, the free energy density becomes:
\begin{eqnarray}\label{eq:Fst_virials_correct_compact}
\frac{\beta F_{\rm st}}V & \approx & -\frac12 \, v_b \, e^{\beta\epsilon_p} \, \rho_{\rm st}^2 + \frac12 \, v_b^2 \, e^{2\beta\epsilon_p} \left( 1 - e^{\beta(\epsilon_t - \epsilon_p)} \right) \rho_{\rm st}^3 \nonumber\\
& & + {\mathcal O}(\rho_{\rm st}^4) \, .
\end{eqnarray}
In particular, as anticipated we recover the limit of weak sticker-sticker interactions of the {\it annealed} model (Eq.~\eqref{eq:FreeEnergyAnnealedWeak}), where the pair energy term contributes to the renormalization of the three-body interaction term.

%%%\nocite{*}
%\bibliography{biblio.bib}% Produces the bibliography via BibTeX.
%%%
%\bibliographystyle{unsrt}   % oppure apsrev4-2, plain, ecc.

%apsrev4-2.bst 2019-01-14 (MD) hand-edited version of apsrev4-1.bst
%Control: key (0)
%Control: author (8) initials jnrlst
%Control: editor formatted (1) identically to author
%Control: production of article title (0) allowed
%Control: page (0) single
%Control: year (1) truncated
%Control: production of eprint (0) enabled
%

%%%
\end{document}